\documentclass[12pt,preprint]{aastex}

\begin{document}
\def\Msolar{{\rm M}_{\odot}}

\title{Upper bounds on the low-frequency stochastic gravitational wave
background from pulsar timing observations: current limits and future
prospects}

\author{F. A. Jenet\altaffilmark{1}, G. B. Hobbs\altaffilmark{2},
W. van Straten \altaffilmark{1}, R. N. Manchester\altaffilmark{2},
M. Bailes\altaffilmark{3}, J. P. W. Verbiest\altaffilmark{2,3},
R. T. Edwards\altaffilmark{2}, A. W.  Hotan\altaffilmark{4}, J. M.
Sarkissian\altaffilmark{2}, S. M. Ord\altaffilmark{5}}

\altaffiltext{1}{Center for Gravitational Wave Astronomy, University
of Texas at Brownsville, TX 78520 (merlyn@phys.utb.edu)}
\altaffiltext{2}{Australia Telescope National Facility, CSIRO, PO Box
76, Epping NSW 1710, Australia} 
\altaffiltext{3}{Centre for Astrophysics and Supercomputing,
Swinburne University of Technology, P. O. Box 218, Hawthorn VIC 3122,
Australia}
\altaffiltext{4}{Physics Department, University of Tasmania, Hobart, 
TAS 7001, Australia}
\altaffiltext{5}{School of Physics, University of Sydney, NSW 2006, Australia}

\begin{abstract}
 Using a statistically rigorous analysis method, we place limits on
 the existence of an isotropic stochastic gravitational wave background using
 pulsar timing observations. We consider backgrounds whose
 characteristic strain spectra may be described as a power-law dependence with
 frequency. Such backgrounds include an astrophysical background
 produced by coalescing supermassive black-hole binary systems and
 cosmological backgrounds due to relic gravitational waves and cosmic
 strings.  Using the best available data, we obtain an upper limit on
 the energy density per unit logarithmic frequency interval of
 $\Omega^{\rm SMBH}_g(1/8\mbox{ yr}) h^2 \leq 1.9\times 10^{-8}$ for
 an astrophysical background which is five times more stringent than
 the earlier Kaspi et al. (1994) limit of $1.1\times 10^{-7}$. We also
 provide limits on a background due to relic gravitational waves and
 cosmic strings of $\Omega^{\rm relic}_g(1/8\mbox{ yr}) h^2 \leq
 2.0\times10^{-8}$ and $\Omega^{\rm cs}_g(1/8\mbox{ yr}) h^2 \leq
 1.9\times10^{-8}$ respectively. All of the quoted upper limits
s correspond to a 0.1\% false alarm rate together with a 95\% detection
 rate. We discuss the physical implications of these results and
 highlight the future possibilities of the Parkes Pulsar Timing Array
 project.  We find that our current results can 1) constrain the
 merger rate of supermassive binary black hole systems at high
 red shift, 2) rule out some relationships between the black hole mass
 and the galactic halo mass, 3) constrain the rate of expansion in the
 inflationary era and 4) provide an upper bound on the dimensionless
 tension of a cosmic string background.

\end{abstract}
\section{Introduction}
Pulsar timing observations (see Lorimer \& Kramer 2005\nocite{lk05},
Edwards, Hobbs \& Manchester 2006\nocite{ehm06} for a review of the
techniques) provide a unique opportunity to study low-frequency
($10^{-9}-10^{-7}$ Hz) gravitational waves (GWs) (e.g., Sazhin
1978\nocite{saz78}, Detweiler 1979\nocite{det79}, Bertotti, Carr, \&
Rees 1983\nocite{bcr83}, Foster \& Backer 1990\nocite{fb90}, Kaspi,
Taylor \& Ryba 1994\nocite{ktr94}, Jenet et
al. 2005\nocite{jhlm05}). Sources in this low-frequency band include
binary supermassive black holes, cosmic super-strings, and relic
gravitational waves from the Big Bang (Jaffe \& Backer 2003, Maggiore
2000\nocite{mag00}).

An isotropic stochastic background can be described by its characteristic strain
spectrum $h_c(f)$ which, for most models of interest, may be written
as a power-law dependence on frequency, $f$:
\begin{equation}
h_c(f) = A \left(\frac{f}{{\rm yr}^{-1}}\right)^\alpha.
\label{hc}
\end{equation}

Table~\ref{table1} shows the expected values of $A$
and $\alpha$ for different types of stochastic backgrounds that have
been addressed in the literature. The characteristic strain is related to the one-sided
power spectrum of the induced timing residuals, $P(f)$, as
\begin{equation}
P(f) = \frac{1}{12 \pi^2}\frac{1}{f^3}h_c(f)^2,
\label{resspec}
\end{equation}
and to $\Omega_{gw}(f)$, the energy density of the background per unit
logarithmic frequency interval, as
\begin{equation}
\Omega_{gw}(f) = \frac{2}{3}\frac{\pi^2}{H_0^2} f^2 h_c(f)^2
\label{omega}
\end{equation}
where $H_0$ is the Hubble constant. Note that the one-sided power spectrum, $P(f)$, is defined so that 
\begin{equation}
\int_0^\infty P(f) df = \sigma^2
\end{equation}
where $\sigma^2$ is the variance of the arrival time fluctuations, or
timing residuals, generated by the presence of the GW
background. Since $\sigma^2$ has the physical units of s$^2$, P(f) has
the units of s$^2$/Hz or s$^3$.


 Jenet et al. (2005)\nocite{jhlm05} developed a technique to make a
 definitive detection of a stochastic background of GWs by looking
 for correlations between pulsar observations. It was shown that
 approximately 20 pulsars would need to be observed with a timing
 precision of $\sim$100\,ns over a period of $5$\,years in order to
 make such a detection if the GW background is at the currently
 predicted level (Jaffe \& Backer 2003, Wyithe \& Loeb 2003, Enoki et
 al. 2004, Sesana et al. 2004)\nocite{shmv04}. The Parkes Pulsar Timing Array (PPTA)
 project (Hobbs 2005\nocite{hob05}) is trying to achieve these
 ambitious goals, but the currently available data-sets do not provide
 the required sensitivity for a detection. In this paper,
 we introduce a method to place an upper bound on the power of a
 specified stochastic GW background using observations of multiple
 pulsars. Full technical details of our implementation will be
 published in Hobbs et al. (2006).  Here, this method is applied to
 data (see Section 2) from seven pulsars observed for the PPTA project
 combined with an earlier publicly available data-set.


\begin{table*}
\caption{The expected parameters for predicted stochastic backgrounds}\label{table1}
\begin{tabular}{llll}\hline \hline
Model & A & $\alpha$ & References \\ \hline

Supermassive black holes & $10^{-15} - 10^{-14}$ & $-$2/3 & Jaffe \& Backer (2003)\\
 & & & Wyithe \& Loeb (2003)\\
 & & & Enoki et al. (2004) \\
Relic GWs & $10^{-17} - 10^{-15}$ & $-1$ -- $-0.8$ & Grishchuk (2005) \\
Cosmic String  & $10^{-16} - 10^{-14}$ & $-$7/6  & Maggiore (2000) \\
\hline\end{tabular}
\end{table*}\nocite{jb03}\nocite{wl03a}\nocite{eins04}

 Upper limits have already been placed on the amplitude of any such
 background of GWs. Using eight years of observations for
 PSR~B1855$+$09, Kaspi et al. (1994) obtained a limit of $\Omega_g
 h^2 \le 1.1 \times 10^{-7}$, where $H_0 = 100h$
 km~s$^{-1}$~Mpc$^{-1}$, at the $95\%$ confidence level\footnote{Their
   more stringent constraint of $\Omega_g h^2 \le 6 \times 10^{-8}$
   was obtained when data from PSRs~B1855$+$09 and B1937$+$21 were
   combined. Since the data from PSR~B1937+21 is far from white, we
   believe this limit is artificially low and therefore restrict our
   discussion to the PSR~B1855+09 data only.} for the case when $\alpha =
   -1$ (i.e. $\Omega_{gw}$ is independent of frequency). This work was
   continued by Lommen (2002)\nocite{lom02} who used 17\,yrs of
   observations to obtain $\Omega_g h^2 < 2\times10^{-9}$.  However,
   the statistical method used for both of these analyses has been
   criticized in the literature (see, for instance, Thorsett \& Dewey
   1996\nocite{td96}, McHugh et al. 1996\nocite{mzvl96}, Damour \&
   Vilenkin 2005). In this paper, we develop a frequentist technique,
   similar to that used by the LIGO science collaboration (Abbott et
   al. 2006\nocite{aaa+06}), to place an upper bound on $A$, given
   $\alpha$. The technique makes use of a statistic, $\Upsilon$,
   defined below, which is sensitive to red noise in the pulsar timing
   residual data. Upper bounds on $A$ are determined using $\Upsilon$
   together with a specified false alarm rate, ${\cal P}_f$, and
   detection rate, ${\cal P}_d$.  Monte-Carlo simulations are used to
   determine these probabilities by generating pulsar pulse
   times-of-arrival consistent with a GW background. All of the
   upper limits quoted in this paper correspond to ${\cal P}_f = 0.1\%$
   and ${\cal P}_d = 95\%$.

\section{Observations}
We expect the isotropic background to generate timing residuals with a
``red'' spectrum: a spectrum with excess power at low frequencies or,
equivalently, long time scale correlations in the
residuals. Therefore, we have restricted our analysis to those pulsars
having formally white spectra: a spectrum with statistically equal
power at all frequencies or no correlations in the residuals. This
allows us to put the best upper limit on the background by bounding
the level of any red process in those data sets. Three separate tests
were used in order to determine the statistical properties of the data
and to select data-sets that are statistically white. First, the
normalized Lomb-Scargle periodogram was calculated for each residual
time series. No significant peaks were seen in any of the data
used. Second, the variance of the residuals was shown to decrease as
$1/n$ where $n$ is the number of adjacent time samples averaged
together. If the data were correlated the variance would not scale as
$1/n$. Third, no significant structures were seen in the polynomial
spectrum (defined below) for each individual spectrum or in the
averaged spectrum. Note that the publicly available data-set for
PSR~B1937+21 (Kaspi et al. 1994) was not used in our analysis since
its timing residuals do not pass these three tests.

We made use of the following data-sets which passed the tests: 1)
observations of PSR~B1855+09 (also known as PSR~J1857+0943) from the
Arecibo radio telescope that are publicly available (Kaspi et
al. 1994), 2) observations for PSRs~B1855+09, J0437$-$4715,
J1024$-$0719, J1713+0747, J1744$-$1134, J1909$-$3744 and B1937+21
(J1939+2134) using the Parkes radio telescope and reported by Hotan et
al. (2006)\nocite{hbo06} and 3) recent observations of all of these
pulsars made as part of the PPTA and related Swinburne timing
projects. The Kaspi et al. (1994) data-set was obtained at
$\sim$1400\,MHz over a period of 8\,yrs.  The PPTA observations, which
commenced in February 2004, include $\sim$20 millisecond pulsars and
use the 10/50-cm dual-frequency receiver and a 20-cm receiver at the
Parkes radio telescope.  Each pulsar is typically observed at all three
frequencies in sessions at intervals of 2 -- 3 weeks.  The results
used here were obtained using a correlator with 2-bit sampling capable
of bandwidths up to 1\,GHz and a digital filterbank system with 8-bit
sampling of a 256\,MHz bandwidth.  The PPTA observations and the
earlier Hotan et al. (2006) data-sets also used the Caltech Parkes
Swinburne Recorder 2 (CPSR2 -- see Hotan et al. 2006), a baseband
recorder that coherently dedisperses two observing bands of 64 MHz
bandwidth, centered on 1341 and 1405 MHz for observations at 20 cm and
around 3100 MHz and 685 MHz for (simultaneous) observations with the
coaxial 10/50 cm receiver. Full details of the PPTA project will be
presented in a forthcoming paper; up-to-date information can be
obtained from our
web-site\footnote{\url{http://www.atnf.csiro.au/research/pulsar/ppta}}.
Unfortunately, our stringent requirements on the ``whiteness'' of the
timing residuals has restricted the use of some of our nominally
best-timing pulsars.  For instance, even though a 10-yr data span is
available for PSR~J0437$-$4715, the full-length data-set is
significantly affected by calibration and hardware-induced artifacts
as well as other unknown sources of timing noise.

\begin{figure}
\includegraphics[width=4in,angle=-90]{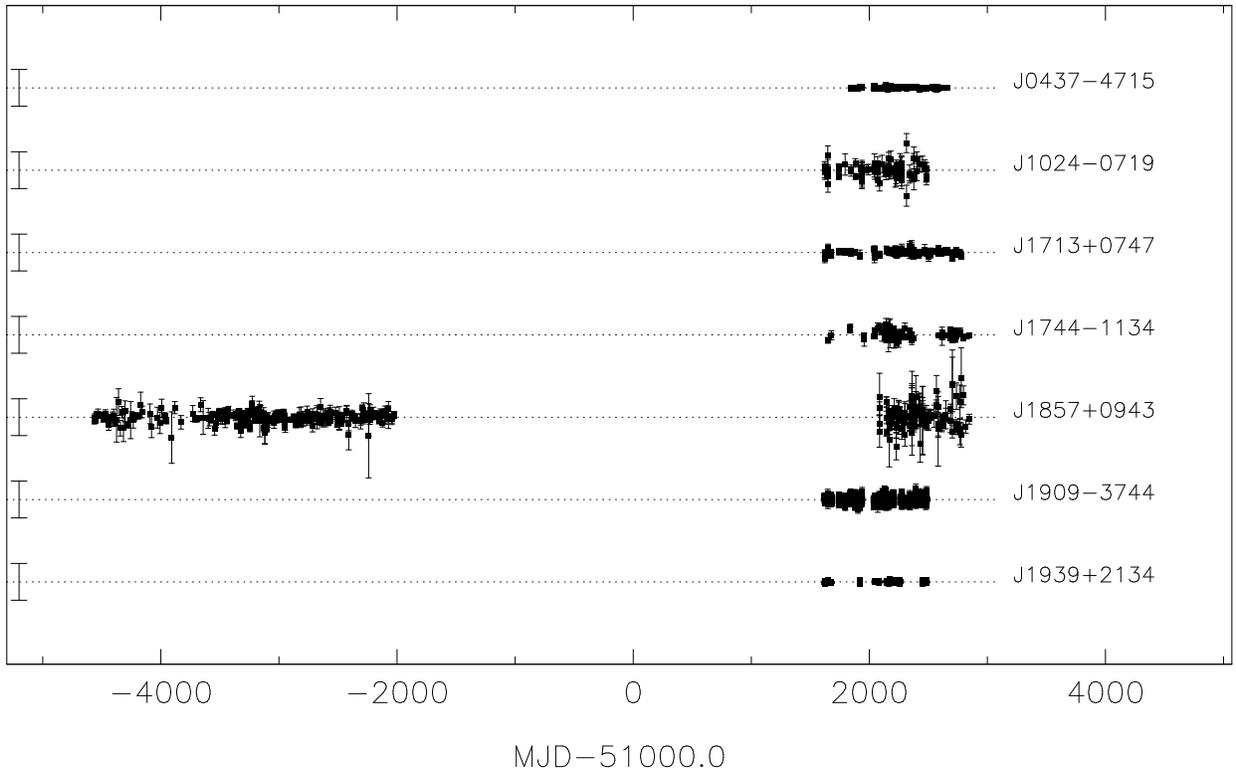}
\caption{Pulsar timing residuals.  The length of the vertical line on
the left hand edge represents 10$\mu s$.}\label{fg:res}
\end{figure}

 A listing of the pulsars observed, the observation span, number of
 points and weighted rms timing residual after fitting for the
 pulsars' pulse frequency and its first derivative, astrometric and
 binary parameters are presented in Table~\ref{table2}. Arbitrary
 offsets have been subtracted between data-sets obtained with
 different instrumentation. Combining these data-sets provides us with
 data spans of $\sim$20\,yr for PSR~B1855+09 and $\sim 2-4$\,years for
 the remaining pulsars.  The final timing residuals are plotted in
 Fig.~\ref{fg:res}.


\begin{table}
 \begin{center}
 \caption{Pulsar observations used for this analysis}\label{table2}
 \begin{tabular}{lcllll} \hline \hline
 Pulsar       & Telescope & Span & N & rms residual \\
              &           & (d)  &   & ($\mu s$) \\ \hline
 J0437$-$4715 & Parkes    & 815  & 233 & 0.12 \\
 J1024$-$0719 & Parkes    & 861  & 92  & 1.10 \\
 J1713+0747 & Parkes    & 1156 & 168  & 0.23 \\ 
 J1744$-$1134 & Parkes    & 1198 & 101  & 0.52 \\
 J1857+0943 & Arecibo/Parkes & 7410  & 398 & 1.12 \\
 J1909$-$3744 & Parkes  & 866 & 2859 & 0.29 \\
 J1939+2134 & Parkes  & 862 & 231 &  0.21 \\
 \hline \end{tabular}
 \end{center}
\end{table}
\section{New upper bounds on the stochastic backgrounds}



 The goal here is to use the measured timing residuals from multiple
 pulsars in order to determine the smallest value of $A$ that can be
 detected for a given $\alpha$ as defined by equation \ref{hc}. This
 is done in a three-step process. First, a detection algorithm is
 defined that is sensitive to the presence of the background. Second,
 this algorithm is tuned so that in the absence of a signal
 (i.e. $A=0$), the probability of the detection scheme falsely
 detecting the background is set at ${\cal P}_f$, known as the false alarm
 rate. Lastly, for the given detection scheme and false alarm rate,
 the upper bound, $A_{\rm upper}$, is chosen so that the probability
 of detecting a background with $A = A_{\rm upper}$ is
 ${\cal P}_d$. For this paper, the false alarm rate is set to $0.1\%$
 while the upper-bound detection rate is set to $95\%$.

 Since all current models of the background predict that the induced
 timing residuals will be ``red'' (the spectrum increases at lower
 frequencies), the detection scheme employed here is defined to be
 sensitive to a red spectrum. The existence of a red spectrum in the
 timing residuals is therefore necessary, but not sufficient, evidence
 for the existence of a GW background. Hence, we can use a
 statistic sensitive to a red spectrum in order to place an upper
 bound on the amplitude of the characteristic strain spectrum. Since
 the data-sets are irregularly sampled and cover different time spans,
 a spectrum based on orthogonal polynomials is employed. Each pulsar
 data-set consists of $n_p$ measured residuals, $x_p(i)$, a time tag
 $t_p(i)$, and an uncertainty $\sigma_p(i)$, where $i$ is the data
 sample index and $p$ is an index referring to a particular pulsar. The
 time tags are scaled so that normalized time tags, $\tau_p(i)$, run
 from $-1$ to $1$.
 These $\tau_p(i)$ values are used in a weighted Gram-Schmidt
 orthogonalization procedure to determine a set of orthonormal
 polynomials $j_p^l(i)$ defined from
\begin{eqnarray}
\sum_{i=0}^{n_p-1} \frac{j_p^l(i) j_p^k(i)}{\sigma_p^2(i)} = \delta_{lk}
\end{eqnarray}
where $j_p^l(i)$ is the $l$'th order polynomial evaluated at
$\tau_p(i)$ and $\delta_{lk}$ is the standard Kronecker delta
function. Note the highest power of $t$ in $j_p^l(i)$ is $l$. For the
case where $\tau$ is continuous and $\sigma_p^2(i)=1$, the above sum
becomes an integral and $j_p^l(i)$ become the familiar Legendre
polynomials. The following coefficients are calculated using the
orthonormal polynomials, $j_p^l(i)$, and the timing residuals
$x_p(i)$:
\begin{equation}
C^l_p = \sum_{i=0}^{n_p-1}\frac{j_p^l(i)x_p(i)}{\sigma_p^2(i)}.
\end{equation}
The pulsar average polynomial spectrum is given by 
\begin{equation}
P_l = \sum_p \frac{(C^l_p)^2}{v_p}
\end{equation}
where the weighted variance, $v_p$, is defined as $\frac{1}{n_p}
\sum_{i=0}^{n_p-1}(x_p(i)-\bar{x}_p)^2/\sigma_p^2(i)$ and $\bar{x}$ is
the mean of $x$. Since the stochastic background is red, $P_l$ will be
large for low values of $l$ if the background significantly influenced
the residuals. Hence, $\Upsilon = \sum_{l=0}^{l=7}P_l$ may be used as
a statistic to detect the background. An upper limit of seven is used
since $95\%$ of the power is contained in the first seven polynomials
for the case of $\alpha=-2/3$. The background will be ``detected'' if
$\Upsilon > \Upsilon_0$ where $\Upsilon_0$ is set so that the
false-alarm rate is given by ${\cal P}_f$.


A Monte-Carlo simulation was used to determine $\Upsilon_0$ and
$A_{\rm upper}$. Complete details of the simulation and its
implementation may be found in Hobbs et al. (2006), but a brief
overview is given here. The simulation, undertaken in the pulsar
timing package \textsc{tempo2} (Hobbs, Edwards \& Manchester
2006\nocite{hem06}), generates an ideal TOA data-set (with the same
sampling as the observed data) from a measured set of TOAs and a given
timing model. The fluctuations due to the GW background for a
given $A$ and $\alpha$ are introduced into the TOAs by adding together
10,000 sinusoidal GWs which come from random directions on the sky
and have randomly chosen frequencies in the range $(1/2000 {\rm
  yr})$--$(1/0.5 {\rm d})$. As a test of the simulation, the
ensemble-averaged power spectrum of the simulated residuals was
calculated over a time scale much larger then the longest GW
timescale (i.e. 2000 years) and was shown to be consistent with
equation~\ref{resspec} as expected. The GW residuals are then added to
the ideal TOA data-set for each pulsar. In order to include the
effects of measurement noise, the measured timing residuals are added
back into the data-set, but randomly shuffled. This ensures that the
added noise has the same probability distribution as the actual
measurement noise. In this way, a new set of TOAs are generated that
include both measurement noise and the GW background. Note that
the shuffling procedure is only valid when the data have a white
spectrum. Otherwise, the spectral properties of the original data set
and the shuffled data set will not be the same. This simulated TOA data set
will then be analyzed in exactly the same way as a real data
set. Hence, all the systemic effects which inhibit gravitational wave
detection such as low order polynomial removal, Earth's orbital
motion, annual parallax effects, and orbital companion effects are
appropriately accounted for in the simulation.

To calculate $\Upsilon_0$, the simulation generates 10000 independent
simulated sets of TOAs for each pulsar with $A=0$ (i.e. no GW
background). The statistic $\Upsilon$ is calculated for each of the
10000 trials. Using this set of $\Upsilon$ values together with the
chosen false alarm rate, ${\cal P}_f$, the value of $\Upsilon_0$ can be
determined. Once $\Upsilon_0$ is chosen, the simulation is used to
generate TOA data-sets that include the effects of GWs. For a
given value of $A$, the probability of detection is determined using
$\Upsilon$ and $\Upsilon_0$. $A_{\rm upper}$ is chosen to be that
value of $A$ when the probability of detection is equal to
${\cal P}_d$.

Note that the effects of unknown time offsets (``jumps'') in the
data-sets are included in the calculation of both $\Upsilon_0$ and
$A_{\rm upper}$ using this technique since \textsc{tempo2} fits for
these offsets in the TOA data-set after the GW background has been
added. Since we are using \textsc{tempo2} to analyze the data, the
effects of all the fitting procedures are being taken into account.

\section{Results}
\begin{figure}
\includegraphics[width=4in,angle=-90]{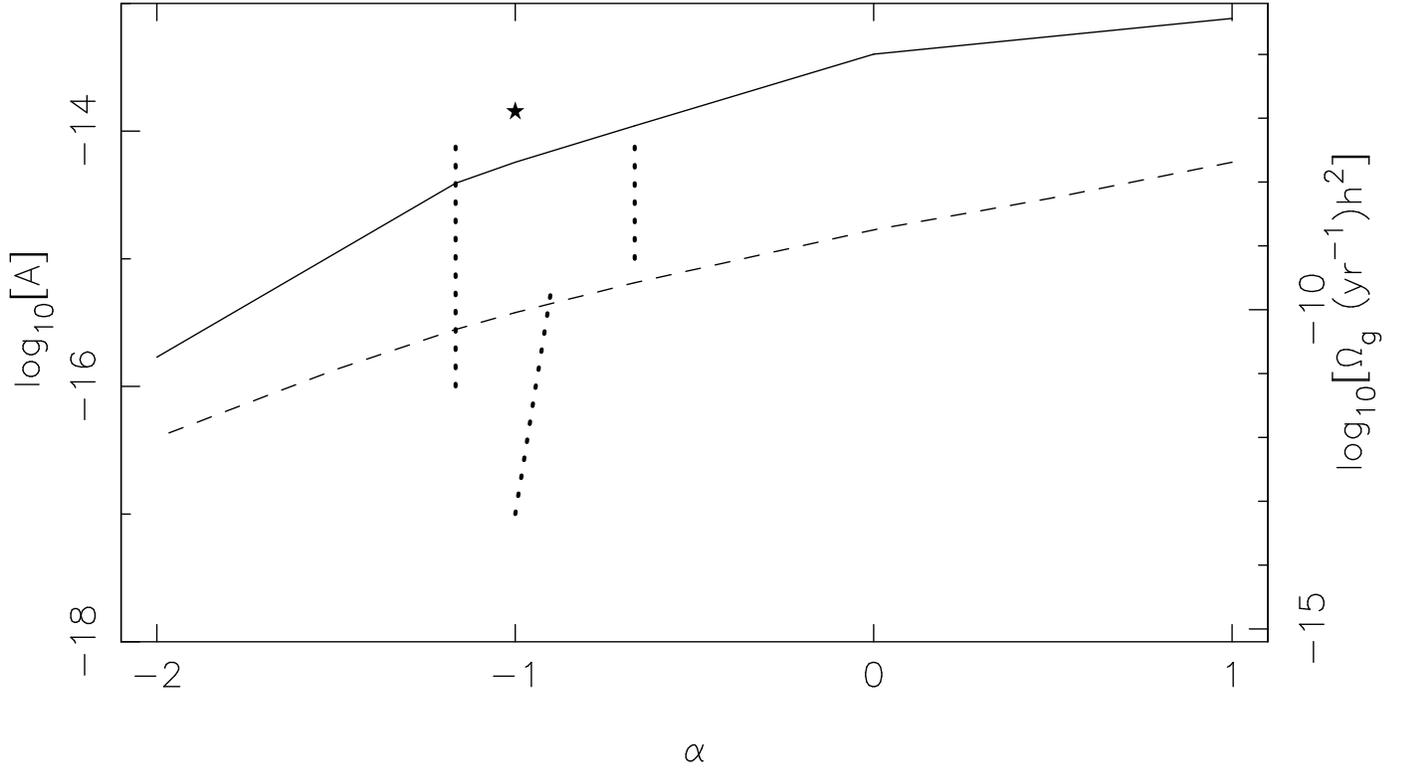}
\caption{\label{Avsa} Minimum detectable A (or $\Omega_{gw}(1 {\rm
  yr}^{-1})h^2$ - right axis) versus $\alpha$ for our current limits
  (solid line) and potential future limits from the PPTA (dashed
  line).  The star symbol indicates the limit obtainable using the
  Kaspi et al. (1994) observations of PSR~B1855+09.  From left to right the
  near-vertical dotted lines indicate the expected range of amplitudes
  for the cosmic strings, relic GW and supermassive black hole
  background respectively.}
\end{figure}
Using the pulsar data-sets described above, the $95\%$ detection rate
upper bound with a false alarm rate of $0.1\%$ is given in Table
\ref{table3} for different values of $\alpha$. The relationship
between $A$ and $\alpha$ is shown in Fig.~\ref{Avsa}. These upper
bounds on $A$ can be converted to an upper bound on the normalized
GW energy-density per unit logarithmic frequency interval,
$\Omega_{gw}(f)$ using equations~\ref{hc} and \ref{omega}. Our limits
on $\Omega_{gw}(1 {\rm yr}^{-1})$ are indicated on the right-hand axis of
Fig.~\ref{Avsa}.

We can compare our results to the previously published limit of Kaspi
et al. (1994) who obtained that $\Omega_{gw}(1/8\mbox{ yr})h^2 < 1.1
\times 10^{-7}$ (star symbol in Fig.~\ref{Avsa}).  Using the same
data-set as Kaspi et al. (1994) our method provides a similar limit of
$\Omega_{gw}(1/8\mbox{ yr})h^2 < 1.3 \times 10^{-7}$. Combining this
data-set with our more recent observations provides a more stringent
limit of $\Omega_{gw}(1/8\mbox{ yr})h^2 < 1.9 \times 10^{-8}$.

The most stringent limit reported to date was obtained by Lommen
(2002). Unfortunately, these observations are not publicly
available.  In order to compare our technique we use the original
PSR~B1855+09, Kaspi et al. (1994) data-set along with two simulated
white data-sets that realistically model the NRAO 140-foot and Arecibo
observations which form the remainder of the Lommen (2002) data (we
simulate 60 observations with an rms residual of $5\mu s$ between MJDs
47800 and 51360 for the 140-foot telescope and a further 60
observations with an rms residual of $1\mu s$ between MJDs~50783 and
52609 for the most recent Arecibo data). As we simulate neither 
systematic effects nor timing noise our limit will be more stringent
than could be obtainable using the real data-set.  For $\alpha =
-2/3$, we obtained that $A \leq 9 \times 10^{-15}$ corresponding to
$\Omega_{gw}(1/17\mbox{ yr})h^2 = 8 \times 10^{-9}$.  This limit is a
factor of four less stringent than that reported by Lommen (2002).

Using simulated data, the upper bounds that can be expected from
future experiments can be determined.  The goal of the PPTA is to time
20 pulsars with an rms timing residual of 100 ns over 5 years. The
dashed line in Fig.~\ref{Avsa} plots $A$ versus $\alpha$ for such a
data-set which could potentially provide a limit on a background of
supermassive black hole systems of $A_{\rm upper} < 6.5\times10^{-16}$
or $\Omega_{gw}(1/8\mbox{ yr})h^2 \le 6.6\times10^{-11}$
(see Table~\ref{table3}).


In Jenet et al. (2005), techniques to use an array of pulsars to
detect a stochastic background of GWs with $\alpha=-2/3$ were
developed.\footnote{Note that $A$ as defined here is larger by a factor
of $\sqrt{3}$ compared to the definition of $A$ used in Jenet et
al. (2005). The definition used here is consistent with Jaffe \&
Backer (2003) and Wyithe \& Lobe (2003).} Given a value for $A_{\rm
upper}$, one can use such techniques to determine the probability of
definitively detecting the GW background using the completed PPTA
data-sets (20 pulsars with an rms timing residual of 100 ns over 5
years) if $A$ were equal to $A_{\rm upper}$. In terms of the
parameter, $S$, defined in Jenet et al. (2005), a significant
detection would occur if $S>3.1$. This corresponds to a $0.001$ false
alarm rate. For the case of $\alpha=-2/3$, the expected value of $S$
(assuming ideal whitening) is about 4.1 for $A=A_{\rm upper}$. Since
the probability distribution of $S$ is approximately Gaussian, the
probability of $S>3.1$ when $\langle S \rangle=4.1$ is $85 \%$.  Hence,
the GW background would be detected $85 \%$ of the time. For the
case of 10 years of observations, the detection rate increases to over
$99\%$ of the time.

%


\begin{table}
\caption{Current and potential future limits on the stochastic
  gravitational-wave background}\label{table3}
\begin{tabular}{ccccc} \hline \hline
$\alpha$ & A & $\Omega_{gw}(1/1\mbox{ yr}) h^2$  & $\Omega_{gw}(1/8\mbox{ yr}) h^2$ &
  $\Omega_{gw}(1/20\mbox{ yr}) h^2$ \\ \hline
-2/3 & $1.1 \times 10^{-14}$ & $7.6 \times 10^{-8}$ & $1.9\times 10^{-8}$ & $1.0\times 10^{-8}$ \\
-1   & $5.7 \times 10^{-15}$ & $2.0 \times 10^{-8}$ & $2.0\times 10^{-8}$ & $2.0\times 10^{-8}$ \\
-7/6 & $3.9 \times 10^{-15}$ & $9.6 \times 10^{-9}$ & $1.9\times 10^{-8}$ & $2.6\times
  10^{-8}$ \\ 
\hline
-2/3 & $6.5 \times 10^{-16}$ & $2.7 \times 10^{-10}$ & $6.6\times 10^{-11}$ & $3.6\times 10^{-11}$ \\
-1   & $3.8 \times 10^{-16}$ & $9.1 \times 10^{-11}$ & $9.1\times 10^{-11}$ & $9.1\times 10^{-11}$ \\
-7/6 & $2.8 \times 10^{-16}$ & $4.9 \times 10^{-11}$ & $9.9\times 10^{-11}$ & $1.3\times 10^{-10}$ \\ 
\hline \end{tabular}

\tablecomments{The upper half of the table gives limits derived from current
  observations. Limits based on timing 20 pulsars with an rms timing
  residual of 100\,ns over 5\,yr are given in the lower half of
  the table.}
\end{table}


\section{Implications and Discussion}

The upper bound on the stochastic background can be used to probe
several aspects of the Universe. Precisely what is being constrained
depends on the physics of the particular background in question. Here,
both the measured upper bounds using the currently available data and
the expected upper bounds using the full five-year PPTA data-set are
discussed in the context of several GW backgrounds.

\subsection{Supermassive black holes}
A GW background generated by an ensemble of supermassive black
holes distributed throughout the universe has been investigated by
several authors (Jaffe \& Backer 2003, Wyithe \& Loeb 2003, Enoki et
al. 2004). In general, the characteristic strain spectrum for this
background can be written as:
\begin{equation}
h_c(f) = 2.5 10^{-16} h \left(\frac{f}{\mbox{yr}^{-1}}\right)^{-2/3}\left\langle\left(\frac{M_c}{10^{7}\Msolar}\right)^{5/3}\right\rangle^{1/2} \left(\frac{N_0}{\mbox{Mpc}^{-3}}\right)^{1/2} I^{1/2}
\end{equation}       
where 
\begin{equation}
I = \int \frac{N(z)}{N_0} H_0 \frac{a(z)}{\dot{a}(z)} \frac{dz}{(1+z)^{4/3}},
\end{equation}
a(z) is the cosmological scale factor written in terms of red shift,
$z$, $\dot{a}(z)$ is the derivative of $a(z)$ with respect to cosmic
time, $H_0$ is the Hubble constant, the chirp mass $M_c = [M_1 M_2
  (M_1 + M_2)^{-1/3}]^{3/5}$ of a given binary system, $\langle
\rangle$ represents ensemble averaging over all the systems generating
the background, $N(z)$ is the galaxy merger rate as a function of red
shift, and $N_0$ is the present day number density of merged galaxies
that created a black-hole binary system. The values of each of these
physical quantities are currently poorly constrained and each
investigator has chosen a different parameterization. Under the
framework described by Jaffe and Backer (2003), $\langle
M_c^{5/3}\rangle$ and $N_0$ are constrained by observations at the
current epoch to be $\langle M_c\rangle \approx 2.3 \times 10^7
\Msolar$ and $N_0 \approx 1/$\,Mpc$^3$. They parameterized the galaxy
merger rate such that $R(z)$ goes as $(1 + z)^\gamma$. Hence, $I$
depends on $\gamma$. Combining the estimates of $\langle M_c\rangle$
and $N_0$ together with our measured upper bound of $A_{\rm upper} =
1.1\times 10^{-14}$, one finds that $I \leq 3$. Using the full PPTA
after 5 years, one expects $I \leq 0.8$. These constraints together
with the calculations of Jaffe \& Backer (2003) (see Fig.~4 in their
work) constrain $\gamma$. Currently the limit on $\gamma$ is $<2.6$
and with the full PPTA $\gamma < 0.4$. This value is expected to lie
somewhere between $-0.4$ and 2.3 (Carlberg et al. 2000\nocite{ccp+00},
Patton et al. 2002\nocite{ppc+02}). Current PPTA sensitivity
(i.e. using the data presented in this paper) is just above the
expected range, while the full PPTA should be able to place meaningful
constraints on this exponent.

In the Wyithe \& Loeb (2003) work, both $\langle M_c^{5/3}\rangle$ and
$I$ depend strongly on the black-hole
versus galactic-halo mass ($M_{\rm BH}-M_{\rm HM}$) relationship. They
discuss several different scenarios which yield different $M_{\rm
BH}-M_{\rm HM}$ relationships and hence different levels of the
background. For the case of an $M_{\rm BH}-M_{\rm HM}$ relationship
determined by Ferrarese (2002)\nocite{fer02}, the expected value of
$A$ is $2 \times 10^{-15}$. For the $M_{\rm BH}-M_{\rm HM}$
relationship derived from Navarro, Frenk, \& White
(1997)\nocite{nfw97}, $A = 5 \times 10^{-15}$. Using an $M_{\rm
BH}-M_{\rm HM}$ relationship derived from simple considerations of BH
growth by feedback from quasar activity (Wyithe \& Loeb 2003,
Haehnelt, Natarajan \& Rees 1998\nocite{hnr98}, Silk \& Rees
1998\nocite{sr98}), $A \approx 10^{-15}$.  Our measured upper
limit for $\alpha = -2/3$ cannot rule out any of these models.
However, if only a limit is obtained from the full PPTA
observations, it will rule out all of the models described
above.

\subsection{Relic gravitational waves}

A relic gravitational wave background is generated by the interaction
between the large-scale dynamic cosmological metric and quantum
fluctuations of the metric perturbations occurring in the early
universe \citep{gri05}. In the nano-Hz frequency
regime, the background takes the following form:
\begin{equation}
h_c(f) =  h_c(f_h)\left(\frac{f}{H_0}\right)^{\alpha} \left(\frac{a_2}{a_H}\right)^{\frac{1}{2}}
\label{relicgw}
\end{equation}
where $h_c(f_h)$ is the magnitude of the characteristic strain
spectrum at $f=H_0$, $a_H$ is the current value of the cosmological
expansion factor, and $a_2$ is the value of the expansion factor at
the start of the matter-dominated epoch.  Note that this expression is
not valid in the ultra-low frequency regime where $f\approx H_0$. The
notation used here is consistent with \citet{gri05} except for
$\alpha$ which is related to Grishchuk's parameter $\beta$ by $\alpha
= 1 + \beta$. The exponent determines the evolution of the
inflationary epoch which starts the GW amplification process. When
$\alpha = -1$, the scale factor grows exponentially with global cosmic
time. The ratio $a_2/a_H$ is believed to be about
$10^{-4}$. $h_c(f_h)$ is constrained by cosmic microwave background
measurements to be about $10^{-5}$. Using these values and assuming
the validity of the amplification scenario described in Grishchuk
(2005), the upper bound on $A$ may be used to constrain $\alpha$. The
upper bound on $\alpha$ is given by the solution to the following
equation:
\begin{equation}
 h_c(f_h)\left(\frac{1/\mbox{1 yr}}{f_h}\right)^{\alpha} \left(\frac{a_2}{a_H}\right)^{\frac{1}{2}} = A(\alpha).
\end{equation}
The above equation yields $\alpha \leq -0.7$ for the current PPTA and
$\alpha \leq -0.84$ for the full PPTA. Within the theoretical
framework described by \citet{gri05}, if $\alpha$ is larger than
$-0.80$, small scale gravitational waves will effect primordial
nucleosynthesis, while an $\alpha$ less than $-1.0$ will result in an
infinitely large energy density in small scale gravitational
waves. Hence, the full PPTA will be able to place useful constraints
on the relic gravitational wave background. Since $\alpha$ determines
the rate of expansion in the inflationary epoch, it turns out that it
is related to the equation of state of the ``matter'' in that epoch by
\begin{equation}
\frac{P}{\epsilon} = w = \frac{2 - \alpha}{3 \alpha}.
\end{equation}
where $P$ is the pressure and $\epsilon$ the energy density. The full
PPTA will constrain $w$ in the early universe to be greater than
$-1.17$. This would limit inflationary models based on
``quintessence'' and ``phantom energy'' \citep{nog06,pad05}.

\subsection{Cosmic strings}

It has been proposed that oscillating cosmic string loops will produce
gravitational wave radiation (Vilenkin 1981\nocite{vil81}). Recently,
\citet{dv05} discussed the possibility of generating a stochastic
GW background using a network of cosmic superstrings. Using a
semi-analytical approach, they derived the following characteristic
strain spectrum valid in the pulsar timing frequency range (see their
equation 4.8):
\begin{equation}
h_c(f) = 1.6 \times 10^{-14} c ^{1/2} p^{-1/2} \epsilon_{eff}^{-1/6} (h/.65)^{7/6}\left(\frac{G \mu}{10^{-6}}\right)^{1/3} \left(\frac{f}{\mbox{yr}^{-1}}\right)^{-7/6}
\end{equation}
where $\mu$ is the string tension, $G$ is Newton's constant, $c$ is
the average number of cusps per loop oscillation, $p$ is the
reconnection probability, $\epsilon_{eff}$ is a loop length scale
factor, and $h$ is the Hubble constant in units of 100 km/s/Mpc. Note that for the above estimate, $h$ was evaluated at $0.65$ in order to be consistent with \citet{dv05}. The
combination $G \mu$ is the dimensionless string tension which
characterizes the gravitational interaction of the strings. The
predicted string tensions are $10^{-11} \leq G \mu \leq 10^{-6}$
\citep{dv05}. Using the above spectrum together with the measured
upper bound on $h_c$ for $\alpha = -7/6$, an upper bound may be placed
on the dimensionless string tension:
\begin{equation}
G \mu \leq 1.5 \times 10^{-8} c^{-3/2} p^{3/2} \epsilon_{eff} ^{1/2} (h/.65)^{-7/2}.
\end{equation}
As emphasized by \citet{dv05}, the above expression for the upperbound may
be simplified using the fact that both $p$ and $\epsilon_{eff}$ are
less then one and $h$ is expected to be greater than .65:
\begin{equation}
G \mu \leq 1.5 \times 10^{-8} c^{-3/2}.
\end{equation}
Using standard model assumption where $c=1$, the upperbound becomes
$G\mu \leq 1.5 \times 10^{-8}$. This is already limiting the parameter
space of the cosmic string model of Sarangi and Tye (2002).  With the
full PPTA, the limit will become $G \mu \leq 5.36 \times
10^{-12}$. Hence, the full PPTA will either detect GWs from cosmic
strings or rule out most current models.



\section*{Acknowledgments} 

 The authors would like to thank J. Armstrong, T. Damour, and
 L. P. Grishchuck for contributing valuable advice during
 the preparation of this manuscript. Some of the data presented in
 this paper were obtained as part of the Parkes Pulsar Timing Array
 project which is a collaboration between the ATNF, Swinburne
 University and The University of Texas, Brownsville. We thank our
 collaborators on this project.  The Parkes radio telescope is part of
 the Australia Telescope which is funded by the Commonwealth of
 Australia for operation as a National Facility managed by CSIRO. Part
 of this research was supported by NASA under grant NAG5-13396 and the National Science Foundation under grant AST-0545837.
 
\bibliographystyle{apj}
\bibliography{journals,modrefs,psrrefs,crossrefs}

\end{document}